\documentclass{article}

\usepackage{amsmath}
\usepackage{amsfonts}
\usepackage{amssymb}
\usepackage{graphicx}

\begin{document}

\newcommand{\revision}{}

\title{Elastic energy of proteins and the stages of protein folding}

\author{J. Lei\thanks{Zhou Pei-Yuan Center for Applied Mathematics, Tsinghua University,
Beijing 100084, China}, K. Huang\thanks{Physics Department, Massachusetts Institute of Technology, Cambridge,
MA 02139, USA}}

\maketitle

\date{}

\abstract{
We propose a universal elastic energy for proteins, which depends only on
the radius of gyration $R_{g}$ and the residue number $N$. It is constructed
using physical arguments based on the hydrophobic effect and hydrogen
bonding. Adjustable parameters are fitted to data from the computer
simulation of the folding of a set of proteins using the CSAW (conditioned
self-avoiding walk) model. The elastic energy gives rise to scaling
relations of the form $R_{g}\sim N^{\nu }$ in different regions. It shows
three folding stages characterized by the progression with exponents $\nu = 3/5, 3/7, 2/5$,
which we identify as the unfolded stage, pre-globule, and molten globule,
respectively. The pre-globule goes over to the molten globule via a break in behavior akin to a first-order phase transition, which is initiated by a sudden acceleration of
hydrogen bonding.}

\section{Introduction}

The folding of a protein chain in water is driven mainly by the hydrophobic
and hydrogen-bonding interactions \cite{BT}. The general shape of the fold,
or tertiary structure, is designed to have hydrophobic side chains buried in
the interior of the protein, in order to avoid direct contact with water.
However, the side chains are attached to a backbone that welcomes exposure
to water, because of their need for hydrogen bonding. These opposing
tendencies reach mutual accommodation through the formation of secondary
structures --- alpha helices and beta sheets --- which ``use
up'' the hydrogen bonds on the backbone. This involves an intricate interplay
between global geometry and local structure, and each protein seems to
present special problems \cite{Sha:06}. Proteins with different amino-acid
sequences can invoke quite different folding mechanisms \cite{Dag:03}, while
proteins with high sequence similarity can end up with very different folds 
\cite{Ale:05, He:05}. Nevertheless, universal aspects do emerge, if one overlooks
details and concentrate only on a few general properties.

An overall characteristic of the tertiary structure is the radius of
gyration $R_{g}$, the root-mean-square separation between residues. It
serves as a length scale, whose behavior can be studied experimentally \cite{Tak,Uzawa:2004, Kimura:2005}. Stages in the folding process
are characterized by scaling relations of the form $R_{g}\sim N^{\nu }$, where $%
N$ is the number of amino-acid residues. The ``compactness
index'' $\nu $ can be measured through small-angle x-ray scattering \cite{Huang:05}. It can be derived from physical models based on intuitive
reasoning, as pioneered by Flory \cite{Flory:53} and \revision{de Gennes} \cite{Genne:79} in the theory of homopolymers.

The unfolded protein chain, which is akin to a homopolymer, can be described
by Flory's model based on SAW (self-avoiding walk) \cite{Flory:53}. One
assumes a potential energy of the form $a\left( R_{g}^{2}/N\right) +b\left(
N^{2}/R_{g}^{D}\right) ,$ where $a$ and $b$ are temperature-dependent
coefficients, and $D$ is the spatial dimension. The first term is a
stretching energy associated with random walk, for which $R_{g}^{2}$ scales
like $N$. The second term arises from the excluded volume effect, and is
proportional to $N$ times the density. Minimizing the energy with respect to 
$R_{g}$ leads to $\nu =3/\left( D+2\right) $, which gives $\nu =3/5$ for $%
D=3 $.

Hong and Lei \cite{Hong:2008} obtain $\nu =2/5$ for the native state by
statistical analysis of data from PDB (the Protein Data Bank). They also
derive it by generalizing Flory's model. Through detailed arguments, they
generalize Flory's stretching energy to $R_{g}^{2}/N^{\left( 2/\alpha
\right) -1}$, where $\alpha $ is the fractal dimension of the system. This
leads to $\nu =\left( \alpha +2\right) /\left[ \alpha \left( D+2\right) \right] $, which yields $\nu =\left( \alpha +2\right) /\left( 5\alpha
\right) $ for $D=3.$ Taking the fractal dimensions to be $\alpha =1,2,3$ for
polymer in good \revision{solvent}, protein native state, polymer in \revision{poor solvent},
respectively, one obtains $3/5,2/5,1/3$ for the respective indices. The
first case reduces to Flory's SAW model. \revision{For protein in the native state}, the
fractal dimension $\alpha =2$ can be deduced from PDB, and the index $\nu
=2/5$ implies Hooke's law $\left( E-E_{0}\right) \sim R_{g}^{2}$. We note that the native protein is less compact than the collapsed polymer, whose index is $1/3$\cite{Genne:79}. \revision{This is mainly because generally proteins are not fully hydrophobic.}  \revision{The proteins with $70\%$ hydrophobic residues do have indices close to $1/3$\cite{Hong:2008}. Furthermore, native proteins are not well-packed because the secondary structures tend to create interior free volumes\cite{Art:1995, Liang:01}.}

Ptitsyn \cite{Pittsyn:92} has proposed a ``molten globule'' prior to the
native state, in which the tertiary structure has taken shape, together with
a large fraction of the final secondary structures. The main difference from
the native state lies in orientations of side chains, which gradually lock
into native contacts in a time scale of seconds. Thus the molten globule is
almost as compact as the native state, and should have $\nu =2/5$. This is
supported by data from x-ray scattering \cite{Pittsyn:92}. In some proteins
there is evidence for a pre-globule stage \cite{Uversky:96}, which goes over
to the molten globule in a sudden jump. Its observed index is $\nu =0.411\pm
0.016$\cite{Uversky:02}.

In the present investigation, we try to verify these universal features, and
understand them in a unified picture based on the twin actions of the
hydrophobic force and hydrogen bonding. We do this by generalizing Flory's
potential energy to a universal elastic energy for proteins, via analysis and
interpretation of data from computer simulations.

\section{Protein folding in the CSAW model}

We simulate the folding of several proteins using the CSAW (conditioned
self-avoiding walk) model \cite{Huang:2007}. A conformation of the protein
is specified by a set of torsion angles, and side chains are approximated by
hard spheres. The unfolded chain is represented by SAW, and folding comes
about through conditions that create a bias in SAW. The ``self-avoidance'' here means that all atoms on the backbone, as well as the
side chains, are treated as hard spheres with appropriate diameters, and
they are forbidden to overlap one another.

One begins with a SAW of $N$ steps. A trial update is generated by the pivot
algorithm \cite{Li:1995}, and is accepted with a probability given by a
Metropolis MC (Monte-Carlo) algorithm, based on the conformation energy%
\begin{equation}
\epsilon =-g_{1}K_{1}-g_{2}K_{2}.
\end{equation}%
The two terms here correspond to the hydrophobic effect and hydrogen
bonding, respectively. The quantity $K_{1}$ is a total hydrophobic shielding number, which
measures how well hydrophobic side chains are shielded from water by
neighboring protein atoms, and is defined as follows. The \revision{$i$'th} residue has
a hydrophobicity $h_{i}$ given by experiments \cite{KD:82}, and a contact number $k_{i}$,
which is the number of nearest-neighbors of its side chain, in the existing
conformation, and $K_{1}=\sum_{i=1}^{N}h_{i}k_{i}$. Here, we adopt the HP model for the primary sequence, and assign $h_i = 1$ for the hydrophobic residues (L, P, M, W, A, V, F, I), and $h_i = 0$ for the others\cite{GG:1999}. The quantity $K_{2}$ is the
total number of internal hydrogen bonds, which are deemed to be in existence
whenever two legitimate partners have relative positions that conform to the
bond separation and angle. Other effects such as electrostatic and van der
Waals interactions are ignored; but they can be easily included if desired.
The MC updates eventually generate a sequence of conformations that form a
canonical ensemble with respect to the conformation energy $\epsilon $. The
action of water is described entirely through the hydrophobic energy $%
-g_{1}K_{1}$, and the random impacts implicit in random walk.

We assume \revision{that} the parameters $g_{1}$, $g_{2}$ \revision{take} the same \revision{values} for all proteins, which are determined in an earlier study of polyalanine \cite{Lei:2008}. 
The temperature is fixed, and implicit in the values of $g_{1}$, $g_{2}$, but not yet calibrated in the absolute scale.

\begin{table}[h] 
\centering%
{\small
\begin{tabular}{|l|l|l|l|l|}
\hline
Protein name & ID & $N$ & $h$ & Structure \\ 
\hline\hline
Polyalanine & ala20 & $20$ & $1.000$ & 1 alpha helix \\ 
\hline
Antimicrobial LCI & 2b9k & $47$ & $0.383$ & 1 beta sheet \\ 
\hline
Tedamistat & 3ait & $74$ & $0.351$ & 2 sheets \\ 
\hline
Myoglobin & 1mbs & $153$ & $0.379$ & 8 helices \\ 
\hline
\parbox{2cm}{Asparagine\\ synthetase} & 11as & $330$ & $0.397$ & \parbox{1.5cm}{
11 helices\\ 8 sheets}\\ 
\hline
\end{tabular}%
\caption{Proteins simulated. $N$ = number of residues, $h$ = fraction of hydrophobic residues.}}
\label{TableKey}%
\end{table}%

We simulate five proteins, as listed in Table 1. For each protein, folding
starts from an unfolded state created by heating the native state to a high
temperature, and then quenched to a fixed low temperature. The folding
process runs for $4\times 10^{6}$ MC steps, with snapshots taken
every 1000 steps. For each protein, the entire procedure is repeated 30
times to generate 30 folding trajectories, or an ensemble of $1.2\times
10^{4}$ conformations. In the case of myoglobin, folding is extended to a
total of $2.4\times 10^{7}$ MC steps.

\begin{figure}[htbp]
\centering
\includegraphics[width=8cm]{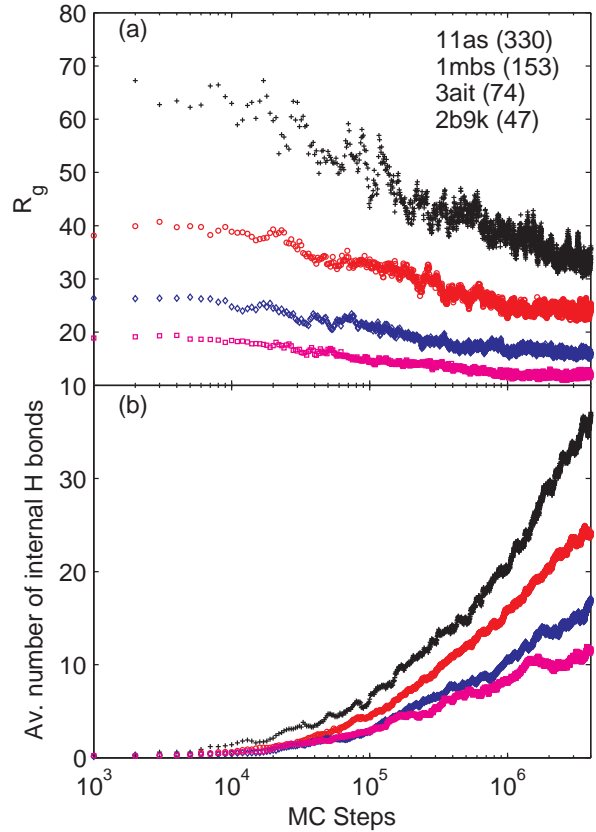}
\caption{Average radius of gyration and number of \revision{internal} hydrogen bonds vs. MC steps from the simulations of four globular proteins, listed with residue numbers in parenthesis, in the order of the curves from top down.}
\label{fig:avepath}
\end{figure}

Fig. \ref{fig:avepath} shows the average radius of gyration and number of \revision{internal} hydrogen bonds as a function of MC steps from the simulations of four globular proteins. From Fig. \ref{fig:avepath}, \revision{we see that} the radius of gyration decreases in the early stage, \revision{and form} a plateau in \revision{a late stage} (except the protein 11as whose simulation has not yet reached the plateau stage). \revision{This is} consistent with experimental observations\cite{Tak,Uzawa:2004, Kimura:2005}. Existence of such stable collapsed states \revision{have} been found in recent AFM experiments\cite{GM:09}. The number of \revision{internal} hydrogen bonds \revision{shows a continuous increase} during the whole process.

\begin{figure}[htbp]
\centering
\includegraphics[width=8cm]{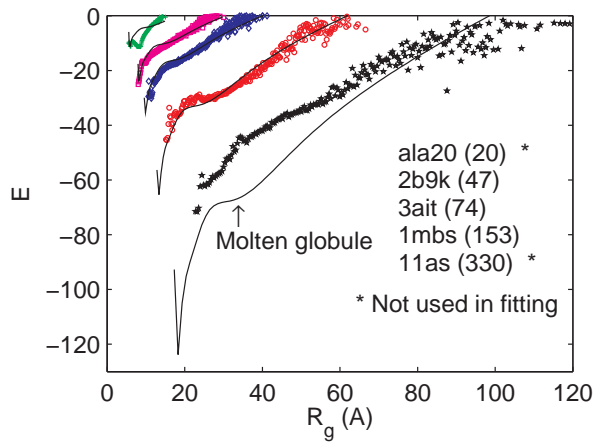}
\caption{Average potential energy vs. radius of gyration, from computer simulations (points) and universal potential \eqref{eq:ene7} (solid curves) of five proteins, listed
with residue numbers in parenthesis, in the order of the curves from top down.}
\label{fig:1}
\end{figure}

We compute an average potential energy $E(R_{g},N)$,$\ $defined as follows.
For each $N$, we take the ensemble average of the model energy $\epsilon $
over all conformations along the folding trajectory that share the same
value of $R_{g}$. Thus, $-\partial E/\partial R_{g}$ gives the
pressure-force experienced by the protein as a function of radius, and can be observed in \revision{force spectroscopy experiments} \cite{Bus:2005, Fer:2004}. In this sense we can
call $E(R_{g},N)$ an ``elastic energy". The simulation
results are shown in Fig.\ref{fig:1}.

\begin{figure*}[htbp]
\centering
\includegraphics[width=14cm]{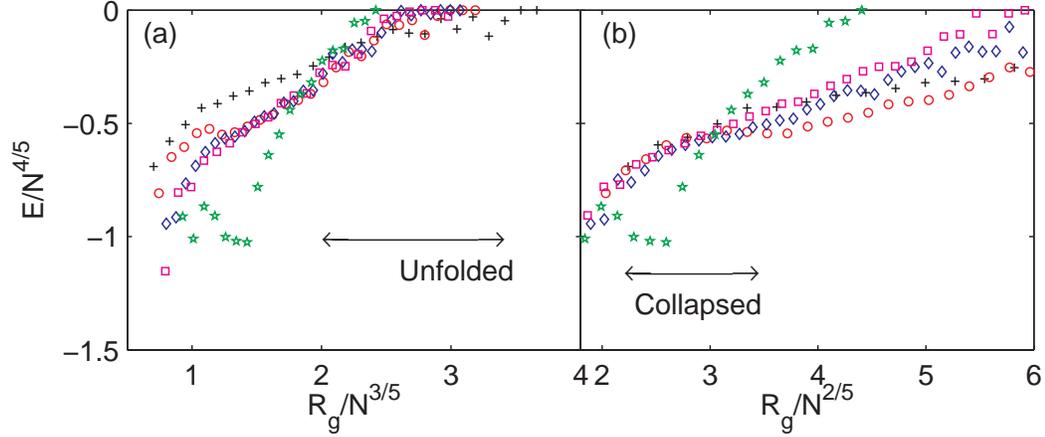}
\caption{\revision{(a) Rescaling the data to exhibit universality in the unfolded stage.
 (b) A different rescaling reveals universality in a collapsed region. The case ala20 (green pentagon) is exceptional, being completely hydrophobic.}}
\label{fig:2}
\end{figure*}

We can display universal features in the data by rescaling the variables (Fig. \ref{fig:2}).
The scales chosen are based on physical pictures, and their validity is to
be judged by goodness of fit. First, we rescale the energy by \revision{a} factor $
N^{4/5}$. This is chosen because in the native state \revision{the scaling laws} $E\sim N^{4/5}$ and $%
R_{g}\sim N^{2/5}$ lead to Hooke's law $E\sim R_{g}^{2}$. On the horizontal
axis of the plot, we rescale the radius by \revision{a} factor $N^{\nu }$ to produce two
separate plots, with $\nu =3/5$ to examine the unfolded region, and $\nu
=2/5 $ to examine the collapsed region. The rescaled graphs are shown in
Fig.\ref{fig:2}(a) for the \revision{3/5} scaling, and Fig.\ref{fig:2}(b) for the \revision{2/5} scaling. As we can
see, the data do exhibit universal behaviors in the respective regions,
except for ala20, which fails to scale in the collapsed region. This
exceptional protein has hydrophobic fraction $h=1$, whereas the others have \revision{average} $
h\approx 0.37 $. This exception, in fact, shows the relevance of hydrophobicity.

\section{A universal elastic energy}

We use physical arguments to suggest an analytical form of a universal
elastic energy, and fit undetermined parameters to simulation data. Let
us start from the unfolded chain, which is indicated in \revision{Fig.\ref{fig:2}(a)}. Assuming a
power law $E/N^{4/5}\sim (R_{g}/N^{3/5})^{p}$, we have $E\sim R_{g}^{p
}N^{(4-3p)/5}$. In the hypothetical limit of a completely extended chain, we
should have $R_{g}\sim N$ and $E\sim N$. This determines $p=1/2,$ and hence 
$E\sim (NR_{g})^{1/2}$. This scaling law describes the hydrophobic energy,
since hydrogen bonding is not significant in this stage. Now we fix the
reference point of energy by taking $E=0$ for a completely extended chain,
which is the convention used in CSAW simulations. Thus we arrive at the
following potential energy in the unfolded region:%
\begin{equation}
E_{1}(R_{g},N)=aN^{4/5}+bN^{1/2}R_{g}^{1/2},
\end{equation}%
where $a$ and $b$ are parameters. As the chain folds, and $R_{g}$ decreases,
the excluded volume effect becomes important. To take this into account, we
add to $E_{1}$ a Flory term $N^{2}/R_{g}^{3}$. Thus, we replace $E_{1}$ with%
\begin{equation}
E_{2}(R_{g},N)=a^{\prime }N^{4/5}+b^{\prime }N^{1/2}R_{g}^{1/2}+c^{\prime
}N^{2}/R_{g}^{3},
\end{equation}%
where $a^{\prime }$, $b^{\prime }$, $c^{\prime }$ are new parameters.

The energy $E_{2}$ has a local minimum corresponding to a metastable state, with scaling law $R_{g}\sim N^{3/7}$, and $E\sim N^{5/7}$. The index $\nu =3/7$ is consistent with the measured value $0.411\pm 0.016$ for the natively unfolded proteins in pre-globule state \cite{Uversky:02}\footnote{Because \revision{of} the hydrogen bonding attraction in \revision{the} pre-globule, the experimental data shows \revision{an} index \revision{closer} to that for the molten globule ($2/5 = 0.4$)\revision{, rather }than the $3/7 = 0.43$.}. 

The universality in the collapsed region (\revision{Fig. \ref{fig:2}(b)}) suggests a power law $E/N^{4/5}\sim (R_g/N^{2/5})^q$, which implies $E\sim R_g^q N^{(4-2q)/5}$. Fitting these forms to the pre-globule state, we find $q=-3$, and hence $E\sim N^2/R_g^3$. This energy involves hydrogen bonding. We \revision{note} that it has the same form as \revision{Flory's excluded-volume energy}, and \revision{thus is} already taken into account in $E_2$.

When a system of hard spheres on a chain collapses, overlaps can only be
avoided through elaborate rearrangements, and the associated thermal energy depends on the intricate structure of the chain, and can only be treated phenomenologically. \revision{We do this by allowing the excluded volume $c^{\prime}$ to depend on $N$ and $R_g$. Since the excluded volume \revision{effect should be} universal for proteins at \revision{the} molten globule \revision{stage}, with \revision{scaling law} $R_g\sim N^{2/5}$, $c^{\prime}$ must \revision{depend} on \revision{$R_g$ through} the scaled radius $\rho =R_{g}/N^{2/5}$.} Thus we arrive at a
universal elastic energy given by%
\begin{equation}
\label{eq:ene7}
E(R_{g},N)=k_{0}N^{4/5}+k_{1}\left( NR_{g}\right) ^{1/2}+N^{4/5}U(\rho ).
\end{equation}%
Here, $U\left( \rho \right) $ is a short-ranged potential, which we expand
in \revision{an} inverse \revision{power series} of its argument:%
\begin{equation}
\label{eq:pow}
U(\rho )=\sum_{n\geq 3}k_{n}\rho ^{-n}.
\end{equation}%

In the power series \eqref{eq:pow}, \revision{the terms with larger exponents} describe  \revision{higher-order} short-range interactions, which depend on \revision{the detailed structure, and are chain specific}. \revision{To obtain a} universal elastic energy with \revision{the least number of} parameters, we limit ourselves to odd powers $3\leq n$ $\leq 15$. \revision{This gives rise} 9 adjustable parameters, which are fitted to simulation data, excluding those from ala20, and 11as. The former is excluded because it is atypical, being 100\% hydrophobic; the latter, because its simulation has not yet reached the pre-globule. With $R_{g}$ measured in angstroms, the parameters are as follows: $k_{0}=-1.50$, and for odd $n$ from $1$ to $15$, $k_{n}=2^n\times$\{$0.43$, $2.53$, $-34.49$, $155.87$, $-326.50$, $350.09$, $-187.19$, $39.59$\}. 

\section{The stages of protein folding}

Graphs of the elastic energy are shown in Fig.\ref{fig:1}. The fits are not ideal for ala20, presumably because the protein is all hydrophobic, nor for 11as, because the simulation in this case is far from complete. 

\begin{figure}[htbp]
\centering
\includegraphics[width=8cm]{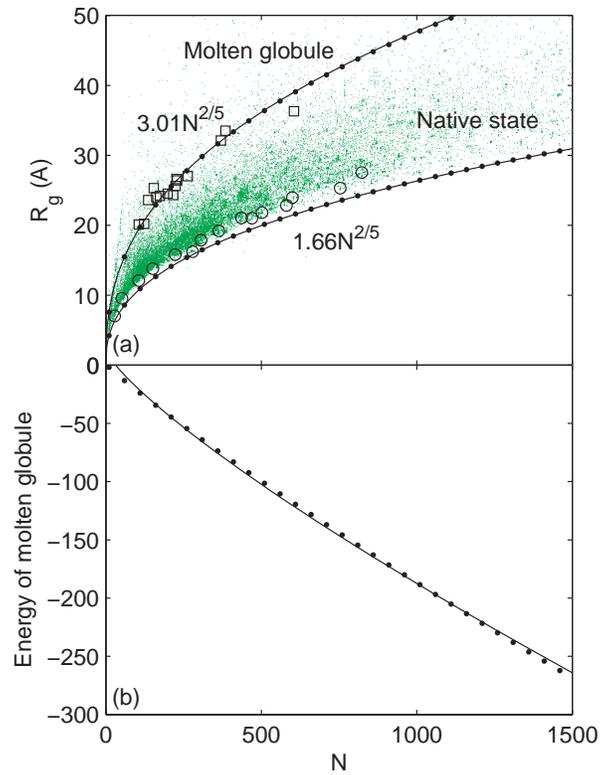}
\caption{(a) Radius of gyration for molten globule: $R_{g}=3.01 N^{2/5}$, and the most compact state: $R_{g}=1.66 N^{2/5}$. Experimental data are also shown, with 37162 proteins in PDB (dots)\cite{Hong:2008}, the most compact proteins from \cite{Art:1995} (circles), and \revision{the} proteins in molten globule state \cite{Tch:2001} (squares). (b) Energy of equilibrium state as a function of $N$: $E = 12.45 - 0.80 N^{4/5}$. The solid points on the curves show that numerical results from the potential\eqref{eq:ene7}.}
\label{fig:3}
\end{figure}

We refer to the curve of 11as for illustration. The region with large $R_{g}$ corresponds to the unfolded stage, which is under pressure to collapse, since $\partial E/\partial R_{g}>0$. The only stable point on the curve is the lowest minimum, which corresponds to the most compact state. \revision{The states with smaller $R_g$ have rapidly increasing energy because of the excluded-volume effect.} There is a flat shoulder corresponding to the molten globule, which can exist in neutral equilibrium. The radii of \revision{the molten globule and the most compact state} are plotted in Fig.\ref{fig:3}(a), with $R_{g} \approx 1.66  N^{2/5}$ for the most compact state, and \revision{$R_{g} \approx 3.01 N^{2/5}$} for the molten globule. We can see from \ref{fig:3}(a) that the above theoretical results agree well with the experimental data. The energy of the molten globule, shown in Fig.\ref{fig:3}(b), obeys $%
\left( E-E_{0}\right) \sim N^{4/5}$, which implies $\left( E-E_{0}\right) \sim R_{g}^{2}$. This agrees with the Hookes-law behavior in the collapsed state. In summary, the three
folding stages --- unfolded, pre-globule, molten globule --- are
characterized by the progression $\nu =3/5,3/7,2/5$ of the compactness index. 

\begin{figure}[htbp]
\centering
\includegraphics[width=8cm]{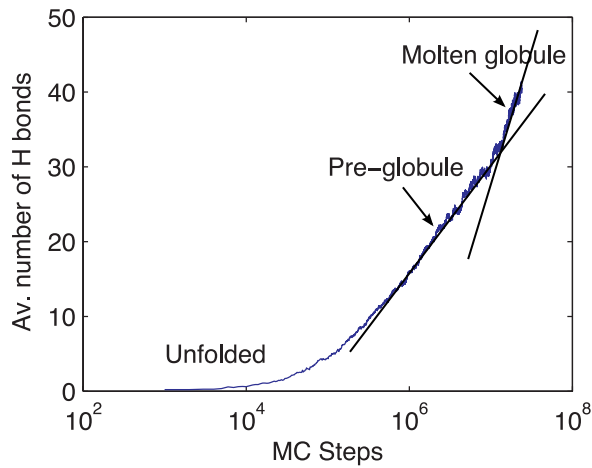}
\caption{Growth of hydrogen bonds in the
simulation of myoglobin. The onset of the transition from pre-globule to
molten globule is marked by a sudden jump in the growth rate.}
\label{fig:4}
\end{figure}

The rate of hydrogen bonding increases in the successive folding stages, as
illustrated in Fig.\ref{fig:4} from the simulation of myoglobin. The transition from
the pre-globule to the molten globule is marked by a sudden acceleration of
hydrogen bonding. This initiates the analog of a first-order phase
transition. Like that in macroscopic matter, the volume is being compressed
at constant temperature, and latent heat is released, since the transition
connects two states of different energy. However, in a small system such as
the protein, there is no clear separation of coexisting phases.

The subsequent evolution from the molten globule to the native state cannot
be described in the present simulation, since it involves the locking of
side chains, and we have approximate them with hard spheres. As we learn
from experimental data, however, $\nu $ should remain unchanged.

\section{Outlook}

The elastic energy here is constructed at a fixed temperature. Work is in
progress to extend it to a range of temperatures, with a view of obtaining a
universal equation of state for proteins. Also under consideration is the
use of the potential in a kinetic equation to compute the lifetimes of the
various stages of protein folding. Although we concentrate on universal
properties here, the CSAW model actually yields results for individual
proteins, which could be analyzed for specificity. The CSAW model is highly
flexible, and amenable to refinements, such as realistic simulation of side
chains, and inclusion of other interactions not yet considered.

This work is supported in part by the National Natural Science Foundation of
China (NSFC10601029).

\end{document}